# Enabling FPGAs for the Masses

Janarbek Matai*, Dustin Richmond*, Dajung Lee†, and Ryan Kastner*
*Computer Science and Engineering, †Electrical and Computer Engineering
University of California, San Diego
{jmatai, drichmond, dal064, kastner}@ucsd.edu

*Abstract*—Implementing an application on a FPGA remains a difficult, non-intuitive task that often requires hardware design expertise in a hardware description language (HDL). High-level synthesis (HLS) raises the design abstraction from HDL to languages such as C/C++/Scala/Java. Despite this, in order to get a good quality of result (QoR), a designer must carefully craft the HLS code. In other words, HLS designers must implement the application using an abstract language in a manner that generates an efficient micro-architecture; we call this process writing *restructured code*. This reduces the benefits of implementing the application at a higher level of abstraction and limits the impact of HLS by requiring explicit knowledge of the underlying hardware architecture. Developers must know how to write code that reflects low level implementation details of the application at hand as it is interpreted by HLS tools. As a result, FPGA design still largely remains job of either hardware engineers or expert HLS designers. In this work, we aim to take a step towards making HLS tools useful for a broader set of programmers. To do this, we study methodologies of restructuring software code for HLS tools; we provide examples of designing different kernels in state-of-the art HLS tools; and we present a list of challenges for developing a hardware programming model for software programmers.

## I. Introduction

FPGAs are a viable platform for accelerating computationally complex applications from a variety of domains such as: computer vision, digital signal processing, and networking. Despite these benefits FPGA designers remain a small set of domain experts with significant knowledge in hardware description languages and underlying architectures. These, along with the required domain specific expertise, are the main road-blocks to new programmers wishing to develop FPGA systems.

Currently, implementing an application on an FPGA is split across two methodologies. The traditional way is using a hardware description languages (HDL) such as Verilog and VHDL. These languages largely operate at the register transfer level, and require intimate knowledge of micro-architectural design techniques. This low level style of coding requires significant hardware design expertise and a long development cycles - often orders of magnitude longer than software development cycles [1]. These characteristics make HDLs an unsuitable method for broadening the number of programmers who can implement an application on an FPGA.

The second design methodology uses High-Level Synthesis (HLS) tools. HLS tools allow designers to use high level languages such as C/C++/Scala/Java. Recently, several HLS tools have emerged from industry [2], [3] and academia [4]–[6]. Most HLS tools accept "software-like" source code and inline optimization directives as inputs. Optimization directives inform the HLS tool about how to optimize parts of the source code. In general, all HLS tools have similar optimization directives with slightly different names. For example, *pipeline* is a common directive in HLS tools. This directive can be given as a *pragma* and performs instruction level pipelining across loops or functions. In general, HLS tools promise to increase the accessibility of designing on FPGAs to a broader number of designers. However, in order to get a good quality of result (QoR) from HLS, the designers must write a *restructured code*, i.e., code that often represents the eccentricities of the toolchain and requires deep understanding of micro-architectural constructs. As noted in previous studies, "standard", off-the-shelf C code typically yields very poor QoR that are orders of magnitude slower than CPU designs, even with HLS optimizations [7], [8].

*Restructured code* is HLS code that is written to target a specific efficient hardware implementation suitable to the FPGA architecture and differs greatly from an initial software implementation. Recent studies suggest that restructuring input code is an essential step to generate an efficient FPGA design [3], [7]–[11]. This means that in order to get efficient designs from HLS tools, the users must write restructured code with the underlying hardware architecture in mind. Therefore, writing restructured code requires hardware design expertise, and domain specific knowledge in addition to software programming skills. The difficulty of writing restructured code limits the FPGA application development to a select number of designers. In other words, if we wish to increase open up FPGA to a larger space of programmers, we must make HLS tools more friendly to programmers that do not have substantial amount of hardware design knowledge.

In the remainder of this paper, we discuss these problems in more detail and propose a solution based on best practices of using HLS. The specific contributions of this paper include:

1) A study on the importance of restructuring code to obtain FPGA designs with good QOR.
2) Two common code restructuring techniques and their impact on final FPGA design QoR for regular and irregular kernels.
3) A list of challenges and possible solutions for developing a novel tool flow that opens up FPGA design to a broader set of programmers.

The remainder of this paper is organized as follows: Section 2 briefly introduces our approach. Section 3 applies these techniques to two application areas. Section 4 describes the performance of these applications. Section 5 presents a list of challenges and possible solutions for developing a programming model for software programmers. Related work and conclusion are presented in Section 6 and Section 7.

## II. Current Approach

Recent state-of-the-art HLS tools reduce the effort reqired to design FPGA applications. These tools eliminate the need to design an architecture that specifies every excruciating detail of what occurs on a cycle by cycle basis. In addition, HLS languages substantially reduce the amount of code that must be written for an application, and in theory enable any software designer to implement an application on an FPGA.

Unfortunately, even a highly optimized "software" implementation of an application will very likely not translate into an optimized "hardware" implementation. Eking out the maximum amount of performance requires significant knowledge of the optimizations that the HLS tool can perform, and basic theories on how the tool implements the design at an architectural level. This is not to say that a programmer needs to know every exact detail of the scheduling, binding, and allocation algorithms in the tool; however, understanding this general process provides great insight that can lead to increased performance. This domain specific knowledge limits the number of programmers that can build "optimized" designs on an FPGA.

In this paper, we address the question: *can a programmer with limited hardware design experience write HLS code that generates efficient FPGA designs?* Evidence indicates that this is not currently possible [3], [7]–[11].

To answer the question above, we present detailed code restructuring methods in HLS code in Section III. We posit that it is possible to generalize code restructuring techniques across common kernels. This idea is inspired by the observations that many applications share core computational kernels, and they can be implemented efficiently on an FPGA with limited insight by the programmer. In the remainder of this paper, we dive deeper into this question and try to determine some necessary steps to ease FPGA development for the masses.

Our work focuses on two different core kernels: *regular kernels* and *irregular kernels*. *Regular kernels* are kernels where the loop bounds are defined and have a direct memory access (e.g., $i$ is an index of A[i], and $for$ loop which has defined upper/lower bounds). *Irregular kernels* are kernels where the loop bounds are dynamic such as loop checking if some error threshold is met such as $while(error < 0.5)$, or kernel has indirect memory access such as A[B[i]] where B[i] is an another array. Previous researchers have shown that *irregular* programs are more difficult to optimize in HLS than *regular* programs [12], [13]. We examine the code reconstructing techniques for both programs with detailed examples in Section III.

## III. RESTRUCTURED CODE

In this section, we demonstrate two code reconstructuring techniques for HLS design based on our experience. We will show that these restructuring techniques are not easy to write for software programmers because they require low-level hardware knowledge, yet produce optimal FPGA designs in HLS. We present software code and reconstructed code for each design study and elaborate on required skills and knowledge for the design process.

### A. Huffman Tree Creation

Huffman encoding is a popular lossless data compression algorithm used in several compression engines such GZIP and JPEG. In modern Huffman encoding, a compression engine calculates a bit length of each symbol. Efficient bit length calculation depends on generating Huffman trees for each text. This section covers the restructuring necessary to transform software-optimized C code for Huffman tree creation into restructured code for current HLS tools.

Listing 1 demonstrates software code for Huffman tree creation, and operates as follows: The input *list* is generated from an arbitrary text source and contains a list of symbols, sorted by the frequency of each symbol in the text. During tree creation, two elements with minimum frequencies (lines 2 and 3) are selected to form a new intermediate node as in line 4. The new node is added to the input list maintaining sorted order (line 7). Figure 1 shows a complete Huffman tree where and $F$ and $B$ have the minimum frequencies of 1. We select $F$ and $B$ to make a new node which will be added to the list with a frequency of 2. The above process continues until no element is left in the sorted input list. The result is the Huffman tree, where the bit length of each symbol is calculated by traversing the tree. For example, $F$ and $B$ have bit length of 4 while $A$ has bit length of 2.

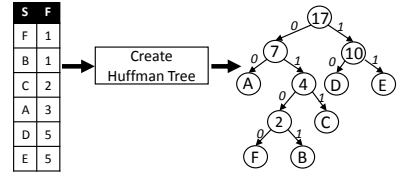

Fig. 1. Huffman Tree: S=Symbol, F=Frequency

*1) Software Code:* Listing 1 presents pseudo code for a subroutine of creating a Huffman tree. This pseudo code is optimized for a CPU application, not for an FPGA design, which leads to drawbacks when it is processed by HLS tools. First, the *while* loop in line 1 is unbounded. Unbounded loop operation prevents hardware pipelining in HLS.

Second, creating a new node (line 4-7) has to re-balance or sort the input list in line 7. Given the code in Listing 1, it is impossible to pipeline the computations in the while loop since the function *InsertToListInSortedOrder* also contains unbounded loops. Finally, the pseudo code is implemented as a recursive function using dynamic memory allocation, which is not possible in current HLS tools.

```
1 while (!isSizeOne(List))
2   Left = extractMin(List[0])
3   Right = extractMin(List[1])
4   Node = newNode(Left->freq + Right->freq)
5   Huffman->CurrentLeft = Left
6   Huffman->CurrentRight = Right
7   InsertToListInSortedOrder(Node)
```

Listing 1. Initial Huffman Tree Creation code as a software code

*2) Restructured HLS Code:* In this section, we present detailed restructured code that creates efficient Huffman tree based on our previous work [8]. Listing 2 presents the pseudo source code that targets the architecture in Figure 2. Figure 2 presents the optimized hardware architecture to create the Huffman tree. This hardware architecture store the intermediate nodes into an hardware array, BRAM.

In the software design, the Huffman tree generation code adds the newly created node to the list maintaining sorter order. The restructured code adds the new node to an empty array and increases its index every time when a new node is created in hardware architecture. This eliminates the computation needed to sort the list on every new node. In addition, the restructured code stores the Huffman tree in a data structure that allows efficient Huffman bit length calculation. After creating initial sub Huffman trees, tree information is stored in three different structures named *ParentAddress*, *Left* and *Right*. Here *Left/Right* store the symbol of left and right children. *ParentAddress* stores the address of the parent of a location where current pointer points. Using these structures, bit length can be calculated in parallel.

In this code, the *HuffmanCreateTree* function has one input, *SF*, and three outputs, *ParentAddress*, *Left* and *Right*. It defines an array in BRAM to store the intermediate nodes in line 5. Its size is $size - 1$ because a Huffman tree with $size$ of leaves has $size - 1$ of intermediate nodes. Lines 6-7 define array indices in the *Left* and *Right* BRAMs. The $while$ loop in line 8 iterates over $SF$ data structure to create the architecture in Figure 2. Lines 11-16 create a left node using the current element of *SF* if $SF.F <= IN.F$ where *SF.F* and *IN.F* are current frequencies of the *SF* and *IN* arrays. Lines 17-23 create a left node using the current element of *IN* (line 20) and saves the index (leftWA) to the *ParentAddress*. In the same way, the lines 25-37 create a right node either using an element from *SF* or an element from *IN*. The $while$ loop in line 38 creates Huffman sub tree if there are intermediate nodes remaining in *IN* array. Both while loops (lines 8 and 38) can be pipelined as shown in the code since they do not contain another loops which are unbounded as in Listing 1.

```
1  void HuffmanCreateTree (
2      SF[size],
3      ParentAddress[size-1],
4      Left[size-1], Right[size-1]){
5  IN[size-1];
6  LeftWA = 0;
7  RightWA = 0; i,k,j=0;
8  while (i<size)
9      #pragma HLS PIPELINE
10     k = k +1
11     if (SF.F <= IN.F){
12         LeftWA = LeftWA + 1
13         Left[LeftWA]=SF[i].S
14         Freq = SF[i].F
15         i = i +1
16     }
17     else {
18         LeftWA = LeftWA + 1
19         Left[LeftWA]= n
20         Freq = IN[i].F
21         ParentAddress[j] = LeftWA
22         j = j +1
23     }
24
25     if (SF.F <= IN.F){
26         RightWA = RightWA + 1
27         Right[RightWA]=SF[i].S;
28         i = i +1
29         IN[k] = SF.F + Freq
30     }
31     else {
32         RightWA = RightWA + 1
33         Right[RightWA]=n;
34         IN[k] = IN.F + Freq
35         ParentAddress[j] = LeftWA
36         j = j +1
37     }
38 while (j < k)
39     #pragma HLS PIPELINE
40     //Create sub trees using IN
41 }
```

Listing 2. Restructured Huffman Tree Creation code for HLS design based on the hardware architecture in Figure 2.

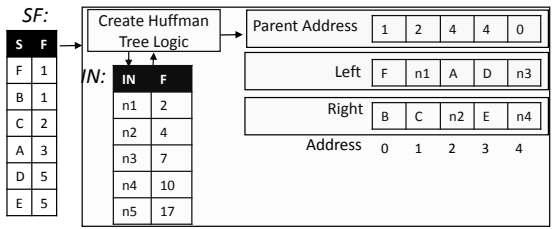

Fig. 2. Hardware architecture of HuffmanCreateTree module: SF is an array storing symbol in S and frequency in F. IN is an array storing symbol in IN field and frequency in F field.

*3) Discussion: HuffmanTreeCreation* code is an example of *irregular kernel* that can be optimized in HLS when we write the code in a restructured way. The restructured code creates the Huffman tree in efficient way despite *while* loops that depend on data and software code that has control dependent statements.

This example shows that the optimization of a HLS design still requires hardware expertise and tool knowledge to write a program as shown in Listing 2. This restructured code generates more efficient hardware design based on the architecture in Figure 2. The software code in Listing 1 is more friendly and intuitive for software engineers, however, a HLS design based on the software version performs poorly due to the nature of the code structure.

*B. Convolution*

Convolution is the most common operation in many image and signal processing applications. Sobel filter is one of example of convolution. This section starts with presenting software code for a Sobel filter. Then we show how to restructure the software Sobel filter in HLS based on [14]. This restructuring is general and can be applied to any convolution kernel.

*1) Software Code:* Sobel filter convolves a given input image with two $3 \times 3$ kernels as described by Equation 1. For each kernel position, it calculates $D_x$ and $D_y$ where $D_x$ and $D_y$ are derivatives for $x$ and $y$ directions. $D_x$ is calculated by multiplying each pixel value of the current $3 \times 3$ input image window of with its corresponding value from $G_x$. Then it takes the sum of the nine multiplications as a value of $D_x$, and repeats the process for $D_y$. The value $D_x + D_y$ is taken as a new value of location at the center of kernel window. Listing 3 shows one of the common ways of writing software Sobel filter in C. The two outer loops iterate over every pixel in the given input image (ignoring boundary conditions for simplicity) and the inner two loops iterate over the values of Sobel filter kernels.

$$G_x = \begin{bmatrix} 1 & 0 & 1 \\ 2 & 0 & 2 \\ -1 & 0 & 1 \end{bmatrix} G_y = \begin{bmatrix} 1 & 2 & 1 \\ 0 & 0 & 0 \\ -1 & -2 & -1 \end{bmatrix} \quad (1)$$

```
1  int image[IMAGE_HEIGHT][IMAGE_WIDTH];
2  for(int i = 0; i < IMAGE_HEIGHT; i++)
3      for(int j=0; j < IMAGE_WIDTH; j++)
4          for(int ro = -1; ro <= 1; ro++)
5              for(int co = -1; co <=1; co++)
6                  D_x += G_x[ro][co] *
                           image[i+ro][j+co] + ...;
7                  D_y += G_y[ro][co] *
                           image[i+ro][j+co] + ...;
8              image[i][j] = D_x + D_y;
```

Listing 3. Software code of Sobel Filter.

*2) Restructured HLS Code:* We must restructure this code to generate an efficient hardware design. A common way of implementing Sobel filter in hardware is through the use of a line buffer and a window buffer. A *line buffer* is a group of memory elements that is capable of storing several lines of an input image. The number of memories, or rows in the line buffer is defined by the height of the kernel. The Sobel kernel has a size of $3 \times 3$ so we use three memories to implement the line buffer. In HLS, we use a 2D array to declare a *line buffer*, i.e., $LineBuffer[3][IMAGE\_WIDTH]$. The *window buffer* stores the values in the current window and it has the same size as the Sobel kernel ($3 \times 3$). We use registers to store window buffer values in order to access them simultaneously in a clock cycle. This code is based on [14].

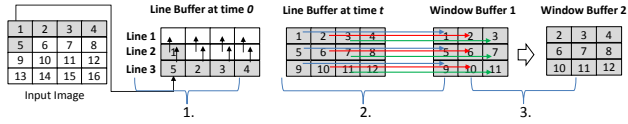

Fig. 3. Line buffer and window buffer example for convolution operation.

Figure 3 demonstrates how the line and window buffers work. For simplicity, we assume that the input image size is $4 \times 4$. The input image is read pixel by pixel into the line buffer. For example, pixel value 1 is copied to the first location, pixel value 2 is copied to the second location of line buffer $line3$, and so on. While copying the input data, line buffers are shifted vertically, and the data from the most upper line buffer is discarded. After time $t$, the $line1$, $line2$ and $line3$ buffers are filled. Since each line buffer is implemented a a separate memory, the first *window buffer 1* can be filled by data from three line buffers in three clock cycles. The next window buffers simply discards the first column and reads the new column from the line buffers in one clock cycle, which enables the data for one Sobel filter operation to be ready every clock cycle.

The restructured HLS code for the architecture in Figure 3 is shown in Listing 4. The code from Lines 7-9 correspond to the first stage. The code from Lines 11-13 correspond to the second stage. In this stage, we design a way to read data from three line buffers to window buffers in parallel. In the last stage, we design hardware for shifting the window buffer by reading a new column from the line buffers. This process is shown in Lines 15-17. After the window buffer is filled with necessary data, we call the $sobel\_filter$ function passing the filled $WindowBuffer$ as an argument. The $sobel\_filter$ function source code is shown in Lines 21-27. The $sobel\_filter$ kernel is computed in one clock cycle using pipelining.

```
int LineBuffer[3][IMAGE_WIDTH];
int WindowBuffer[3][3];

for(int i=0; i<IMAGE_HEIGHT; i++)
 for(int j=0; j<IMAGE_WIDTH; j++)
  #pragma pipeline
  LineBuffer[0][j]=LineBuffer[1][j];
  LineBuffer[1][j]=LineBuffer[2][j];
  LineBuffer[2][j]=image[i][j];

  WindowBuffer[0][0] = LineBuffer[0][j];
  WindowBuffer[1][0] = LineBuffer[1][j];
  WindowBuffer[2][0] = LineBuffer[2][j];

  for(int k = 0; k < 3; k++)
    WindowBuffer[k][2] = WindowBuffer[k][1];
    WindowBuffer[k][1] = WindowBuffer[k][0];

  sobel_filter(WindowBuffer);

sobel_filter(unsigned char window[3][3]){
#pragma pipeline
for(int i=0; i < 3; i++)
  for(int j = 0; j < 3; j++)
    D_x = D_x + (window[i][j] * G_x[i][j]);
    D_y = D_y + (window[i][j] * G_y[i][j]);
    sum = D_x + D_y;
```

Listing 4. Restructured HLS code for the Sobel edge detection.

*3) Discussion:* The source code shown in Listing 4 is the restructured C code for the Sobel filter design. We used the $pipeline$ pragma in line 6 to parallelize everything under line 6. In this implementation, we can fill the window buffer in every clock cycle, and process the window buffer with $sobel\_function$ in next clock cycle. This allows us to achieve a pixel rate of one per clock cycle. The $pipeline$ pragma in (line number 7) instructs HLS to process the code below in every clock cycle. This is the optimal clock cycles achievable by manual design assuming the design processes a new pixel in each clock cycle.

Despite the fact that convolution kernels are regular kernels (automatic compiler optimizations such as polyhedral optimizations are possible) restructured code has its benefits as shown above. The source code in Listing 3 and Listing 4 have the same functionality, but result in very different hardware implementations. We can only optimize Listing 3 by pipelining most inner loop due to memory port limitation on $image$ variable. The memory access pattern of Listing 3 does not allow for the outer loop to be pipelined. The restructured code from Listing 4 achieves the optimal number of clock cycles while the design from Listing 3 needs 67X more clock cycles. Clearly, the code restructuring performed in Listing 4 is necessary to achieve an optimized hardware implementation.

An experienced HLS programmer would write restructured code as in Listings 2 and Listings 4 as this is a standard way to design efficient hardware. This way of thinking architecture and coding is nontrivial task for software programmers. For example, code in Listing 4 requires HLS programmers to think about the hardware architecture at a clock cycle level such as how data moves from input to line buffer, then shifting to window buffer.

## IV. EXPERIMENTAL RESULTS

In this section, we present area and performance results for Huffman tree creation and Sobel kernel convolution. For each kernel, we compare performance/area results of software code and restructured HLS code. Software code is optimized with HLS pragmas (e.g., pipeline) with minimal code changes in order to make them synthesizable. In this work, Vivado HLS 2013.4 is used for the hardware implementations with target device Xilinx Zynq FPGA (xc7z020clg484-1). All results are obtained after place and route. In the following tables, we use *software* design to refer a hardware implementation using the initial software code, and we use *restructured* design to refer the hardware implementation of the restructured code.

### A. Huffman Tree Creation

We created a syntactic data using the LZ77 compression engine with size of 536. Designs using software code and restructured designs are optimized with HLS pragmas on top of them. We performed minimal code restructuring to the pseudo code in Listings 1 in order to make it synthesizable with HLS. Table I shows area and performance results for the Huffman tree creation. Clock cycles are measured from the simulation of the design. Throughput is the number of Huffman tree creations per second. Frequency is in MHz. The first row shows results obtained by implementing the software design. The second row shows the results obtained by implementing the restructured design. The third row (Ratio) is the ratio between components of software designs versus hardware design. Larger (larger than 1) ratio means software design is bad for slices, BRAM and clock cycles. Smaller (smaller than 1) means restructured design is good for throughput and frequency. The frequency of software design is little bit better than restructured design due to limited parallelism in the software design. BRAM usage is decreased from 9 to 2 in the restructured design due to writing a restructured code that is more hardware friendly.

TABLE I. HUFFMAN TREE CREATION.

|  | Area | | Performance | | |
|---|---|---|---|---|---|
|  | Slices | BRAM | Clock Cycles | Throughput | Frequency |
| Software | 295 | 9 | 7889921 | 18 | 145 |
| Restructured | 353 | 2 | 3142 | 39893 | 125 |
| Ratio | 0.83 | 4.5 | 2511 | 4.5e-4 | 1.16 |

*B. Convolution*

Table II shows performance area results for the convolution designs. Software design tends to use 67 times more clock cycles than the restructured design while both designs achieving very similar frequency. As a result, the restructured design has 67 times more throughput than the software design. The software design uses less slices because of limited parallelism and does not use any BRAMs in the logic due to the nature of software code. In restructured code we stored three lines of input image to line buffers which consumes three BRAMs.

TABLE II. CONVOLUTION.

|  | Area | | Performance | | |
|---|---|---|---|---|---|
|  | Slices | BRAM | Clock Cycles | Throughput | Frequency |
| Software | 472 | 0 | 20889601 | 6.2 | 129 |
| Restructured | 627 | 3 | 307200 | 417 | 128 |
| Ratio | 0.7 | 0 | 67 | 0.01 | 1.007 |

V. CHALLENGES

Today's HLS tools are close to overcoming challenges of manual hardware (HDL) design. This is due to result of more than three decades of research. Despite this, HLS tools are still the domain of hardware experts. In Figure 4, we show the overview of a proposed tool chain to allow software developers to use HLS tools more easily. In this section, we discuss some challenges to be solved in order to realize this flow.

*A. Restructured Code Generation*

Most software engineers are not familiar with HLS coding styles presented in Section III because it requires developers to write restructured code targeting a specific implementation and knowledge of the underlying FPGA. One way to make the restructuring easier for software programmers is through the use of *automated compiler techniques* and *domain specific HLS templates*. Automated compiler equipped with automatic parallelization and memory optimization techniques such as Polyhedral models promise to efficiently generate optimized HLS code from software code [15]–[17]. However, automatic parallelization alone is not enough since some kernels require creation of efficient hardware architecture.

We propose domain specific HLS templates ease generation of restructured code for common kernels. Domain specific HLS templates *define an efficient hardware architecture* for certain classes of common kernels that have the same or similar computational patterns. Common kernels with the same or similar computational patterns are very prevalent in real world applications, and some research has done to classify kernels according to computational patterns [18]. Current classification techniques mostly target general parallel programming practices (e.g., multi-core CPU). We can classify frequently used kernels in FPGA applications according to hardware architecture. For example, sliding window is a common architectural pattern shared both by Sobel and Gaussian filters and both map to the same hardware architecture, convolution, which can be implemented as an HLS template.

While there is no universal way to classify/extract kernels according to their computational/communicational patterns, we purpose a solution with three steps:

1) Identify common kernels from applications from a variety of applications
2) Classify these kernels according to their efficient hardware architectures.
3) Make domain specific HLS templates for those kernels based on their class of hardware architecture.

While having domain specific HLS template for every kernel is not possible, having domain specific HLS templates for the common kernels will ease the use of FPGA by software programmers. Further research is needed to identify, classify and making domain specific templates for the most common kernels for HLS tools. These domain specific templates must by tool independent and define underlying hardware architecture in an efficient way. Domain specific HLS templates are incorporated into the design flow to design kernels as shown in Figure 4. The communication between these kernels are discussed in next.

*B. Complex Application Design*

Real world applications are complex, and almost universally contain several computational kernels. Therefore, software programmers must be able to connect and map multiple kernels of an application on an FPGA. The main challenge here is: *What is the best way connect the kernels designed with domain specific templates in HLS to facilitate task level parallelism?* State-of-the-art HLS tools provide interface directives such as *ap_fifo* to specify a port as a FIFO, or *ap_memory* to specify a memory port. However, these kind of low level interface optimizations require hardware domain expertise. In general FPGA systems for applications such as video processing, digital signal processing, wireless systems, and data analysis rely on dataflow streaming architectures [19]. The programming model for dataflow streaming programs differ significantly from traditional processor (both CPU and GPU) implementations for software programmers. An easy programming model is needed to allow software programmers to exploit dataflow streaming architectures. In this work, we propose an approach to provide communication among kernels efficiently according to Pattern Description. *Pattern Description* represents common programming models such as streaming dataflow and bulk synchronous (CUDA programming model) based on common parallel programming patterns such as MapReduce, Partition which are known to software programmers.

*C. Design Space Exploration*

Design space exploration (DSE) with HLS tools is an essential feature for exploring the performance, area, and power tradeoff of different architectures and underlying implementations. Currently, DSE with HLS is usually done manually. Automatic, but efficient DSE for given code is needed to allow HLS users to tailor hardware for their specific needs. Since DSE is a difficult problem because of large search space, blindly optimizing the code shown in Listings 1 and Listings 3 does not produce efficient hardware. In fact doing DSE on the provided code will result excessive run-time for a single run with current state-of-the-art HLS tools. Using restructured domain specific templates as an input to DSE will allow automatic and efficient DSE with HLS. This is due to the fact that restructured domain specific templates efficiently capture the hardware architecture. These hardware architectures are represented by small number of restructured HLS codes. As a result, domain specific templates reduce the size of the search space.

*D. End-to-End System Design*

In addition to the programming model challenges presented above, verification and communication with HLS kernel on a real FPGA is essential. Current methods of verifying an HLS core on a real FPGA involve several tool flows and non-trivial IP cores such as, DDR controllers, PCI Express Interfaces, and DMA engines to access FPGA memory from an external processor. Using the vendor specific tools with correct IP cores remains difficult even for hardware engineers. One way to solve this problem is providing easy to use high level communication framework between FPGA and host CPU (or FPGA to FPGA or FPGA to Network). Open source frameworks such as RIFFA provide a good abstraction for software developers to access communicate with an FPGA [20]. Integrating easy to use frameworks such as RIFFA with HLS tools will allow easy verification of application code for software developers.

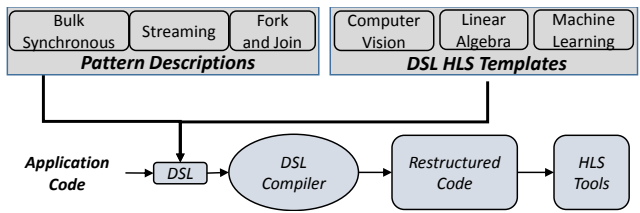

Fig. 4. Design flow for software programmers using HLS templates (restructured code) and parallel programming patterns

## VI. RELATED WORK

There have been several recent case studies on obtaining good QoR for FPGAs using HLS tools [3], [7]–[9]. These works demonstrate that restructuring input C code for HLS is an essential step for generating efficient FPGA design. These works have focused on specific applications and restructured the input C code with promising results.

Recently several works focused on automatically generating code (mostly optimized for memory and data usage based on Polyhedral models) for HLS tools [15]–[17]. Polyhedral models are promising way to automatically generate HLS code for affine-programs. Polyhedral model transformations do not work for non-affine programs such as those with indirect memory accesses (A[B[i]]) or control/data dependent executions [12], [13]. Since polyhedral optimizations are applicable after generating restructured code, above works are orthogonal to our research.

Hardware construction languages such as Chisel [21] and libraries from FPGA vendors (Xilinx OpenCL and Linear algebra) provide a good first step towards making FPGA designs more accessible. While experienced HLS users find these libraries useful, it is difficult for software programmers (still requires substantial hardware expertise) to use them since they still require low level hardware expertise. In order to make them useful for software programmers, higher level of libraries which are easy to use for software programmers are needed.

## VII. CONCLUSION

According to previous studies, results from HLS tools are competitive with manual design techniques using HDLs. However, this requires writing the input code in a way that reflects domain specific hardware knowledge, which we call restructured code. Code restructuring still remains the HLS developer's task and requires hardware expertise. In this work, we presented an approach that promises easier code restructuring easier for software developers. We first presented the importance of code restructuring in HLS. Next, we presented our proposed method based on domain specific HLS templates and Pattern Description. As future work, we aim to automate the process of writing restructured code by creating domain specific languages and pattern descriptions for very large applications which will enable FPGAs for software developers.